\documentclass[12pt,preprint]{aastex}
\usepackage{natbib}
\usepackage{times}

\newcommand{\VLSR}{V_{\rm LSR}}
\newcommand{\FeII}{[\ion{Fe}{2}]}
\newcommand{\SII}{[\ion{S}{2}]}
\newcommand{\OI}{[\ion{O}{1}]}

\newcommand{\HeI}{\ion{He}{1}}

\newcommand{\PaG}{Pa${\gamma}$}
\newcommand{\kms}{km~s$^{-1}$}	

\newcommand{\dotdeg}{\rlap.^{\circ}}

\newcommand{\degree}{^{\circ}}

\shorttitle{\FeII\ EMISSIONS OF UY AURIGE}
\shortauthors{Pyo, et al.}

\begin{document}
\title{\FeII\ EMISSIONS ASSOCIATED WITH THE YOUNG INTERACTING BINARY UY AURIGE \altaffilmark{1}}

\author{T{\sc ae}-S{\sc oo} P{\sc yo}\altaffilmark{1}, M{\sc asahiko} H{\sc ayashi}\altaffilmark{2,3}, 
T{\sc racy} L. B{\sc eck}\altaffilmark{4}, \\ C{\sc hristopher} J. D{\sc avis}\altaffilmark{5}, M{\sc ichihiro} T{\sc akami}\altaffilmark{6}
}

\altaffiltext{0}{Based on observations obtained at the Gemini
  Observatory, which is operated by the Association of Universities
  for Research in Astronomy, Inc., under a cooperative agreement with
  the NSF on behalf of the Gemini partnership: the National Council
  (United Kingdom), the National Research Council (Canada), CONICYT
  (Chile), the Australian Research Council (Australia), CNPq (Brazil),
  and CONICET (Argentina).}
\altaffiltext{1}{Subaru Telescope, National Astronomical Observatory of Japan, 650 North A`oh\=ok\=u Place, Hilo, HI 96720, USA}
\altaffiltext{2}{National Astronomical Observatory of Japan, 2-21-1
  Osawa, Mitaka, Tokyo 181-8588, Japan}
\altaffiltext{3}{School of Mathematical and Physical Science, The
  Graduate University for Advanced Studies (SOKENDAI), Hayama,
  Kanagawa 240-0193, Japan}
\altaffiltext{4}{Space Telescope Science Institute, 3700 San Martin
  Drive, Baltimore, MD 21218, USA}
\altaffiltext{5}{Astrophysics Research Institute, Liverpool John
  Moores University, Liverpool Science Park,
  146 Brownlow Hill, Liverpool L3 5RF, UK }
\altaffiltext{6}{Institute of Astronomy and Astrophysics, Academia
  Sinica, P.O. Box 23-141, Taipei 10617, Taiwan}
\email{pyo@subaru.naoj.org}

\begin{abstract}
We present high resolution 1.06 -- 1.28 $\mu$m spectra toward the interacting
binary UY Aur obtained with GEMINI/NIFS and the AO system Altair.  We
have detected \FeII\ $\lambda$~1.257 $\mu$m and \HeI\ $\lambda$~1.083
$\mu$m lines from both UY Aur A (the primary source) and UY Aur B (the
secondary).  In \FeII\ UY Aur A drives fast and widely opening
outflows with an opening angle of $\sim$~90$\degree$ along a position
angle of $\sim$~40$\degree$, while UY Aur B is associated with a
redshifted knot. The blueshifted and redshifted emissions show 
complicated structure between the primary and secondary.
 The radial velocities of the \FeII\ emission features are
similar for UY Aur A and B: $\sim -$100 \kms\ for the blueshifted
emission and $\sim +$130 \kms\ for the red-shifted component.  The
\HeI\ line profile observed toward UY Aur A comprises a central
emission feature with deep absorptions at both blueshifted and
redshifted velocities.  These absorption features may be explained by
stellar wind models.  The \HeI\ line profile of UY Aur B shows only an
emission feature.
\end{abstract}


\section{INTRODUCTION}\label{sec:intro}
UY Aur is a close binary system composed of classical T Tauri stars
separated by 0$\farcs$89 \citep{Close1998,Duchene1999, Hioki2007}.
\citet{Joy1944} was the first to identify the binarity of UY Aur.  The
secondary source UY Aur B is an infrared companion with a large
extinction; \textit{A}$_V =~$22 - 12 magnitude \citep{Koresko1997}.
The spectral types of the primary and secondary sources are M0 and
M2.5, respectively \citep{HK2003}.  In the optical forbidden emission
lines of \OI\ and \SII , \citet{Hirth1997} identified a redshifted jet
(HH 386) extending over a few arcseconds along a position angle (PA)
of $\sim$~220$\degree$.  They also reported that a blueshifted jet was
evident in 1992 December.  The driving source of the jets was not
clear in their data, however; UY Aur A, B, or both sources.  In their
wide-field survey \citet{McGroarty2004} have since failed to detect the
optical jets beyond what was discovered by Hirth et al.

Molecular hydrogen emission has also been detected from UY Aur, though
only from the secondary source, the infrared companion UY Aur B.  This
emission was interpreted as arising from accretion shocks associated
with the circumstellar disk \citep{Herbst1995}.  The circumbinary disk
around UY Aur is the second such disk that has been resolved and
imaged after the GG Tau A system.  It was resolved for the first time
in a millimeter interferometer survey \citep{Dutrey1994}.
\citet{Duvert1998} found that the $^{13}$CO gas in the circumbinary
disk is in Keplerian rotation and estimated the total mass of the
binary as $\sim$~1.2 M$_\sun$.  Using orbital parameters derived from
observations obtained as far back as 1944, \citet{Close1998} estimated
the total mass of the binary to be 1.61$^{+0.47}_{-0.67}$ $M_\sun$.
The PA of the semi-major axis of the circumbinary disk is
135$\degree$; the inclination angle of the disk is $\sim$~42$\degree
\pm$~3$\degree$ with respect to the line of sight.  High spatial
resolution near infrared imaging with adaptive optics (AO) has also
revealed inner cavity betwen  the circumstellar and circumbinary disks as well as clumpy
structure in the circumbinary disk \citep{Close1998, Hioki2007}.
The inner cavity or gap (i.e. the region with relatively low density) is produced
by a continuous transfer of angular momentum from the binary to the outer circumbinary disk \citep{Artymowicz1991}.
Indeed, \citet{Artymowicz1996} and \citet{Rozyczka1997} showed that
material in the outer disk can penetrate the inner gap and episodically accrete onto the lower mass secondary star, provided the material has sufficiently high viscosity and temperature. 
Simulations of the UY Aur system by \citet{Gunther2002} suggest that the mass accretion rate from the outer
circumbinary disk onto the secondary star might be five times higher and more phase dependent than that to the primary star.
However,  \citet{Hanawa2010} conversely showed that usually the mass accretion rate of the primary is higher than that of the secondary, and in late times the two accretion rates become similar.  
Many simulations produce a gas stream bridge between the circumstellar disks,  as well as an accretion flow toward the individual 
circumstellar disks from the outer circumbinary disk \citep[e.g.][]{Fateeva2011}.
\citet{Hanawa2010} showed that the bridge contains a strong shock front caused by the collision of opposing flows.
Such a bridge structure has been detected in high-resolution coronagraphic H-band imaging of SR24 \citep{Mayama2010}.   

Many young stars are born in binary or multiple systems \citep[see
  e.g. the reviews by][]{Mathieu1994,Zinnecker2001,Duchene2007}.  The
binary frequency for solar type main-sequence stars is about 40 \%$-$60 \%
\citep{Duquennoy1991, Fischer1992}, similar to the frequency
(48.9 \%~$\pm$~5.3 \%) observed for the young stars in the Taurus
molecular cloud \citep{Kohler1998}.  Other nearby star forming regions
show a binary freqency of 9 \%-32 \% \citep[][and references
  therein]{King2012a, King2012b}.  Recent studies suggest that
binarity is common in embedded protostars
\citep{Haisch2004,Connelley2008a,Connelley2008b,Chen2013}.  Jets or
outflows have been observed from a few tens of multiple low-mass young
stars
\citep{Reipurth1993,Reipurth2000,Takami2003,Murphy2008,Mundt2010}.
For single stars, the most plausible launching mechanism is based on
magnetocentrifugal acceleration in a star-disk system: scenarios
include the disk wind \citep{Konigl2000}, X-wind \citep{Shu2000}, and
stellar wind \citep{Matt2005, Matt2008a, Matt2008b} models.  Jets from
a binary system can be explained if the jets emanate from each single
star-disk system \citep[e.g. L1551
  IRS5:][]{Fridlund1998,Rodriguez1998,Itoh2000,Pyo2002,Pyo2005},
although it has been suggested that one of the binary jets could be
destroyed or engulfed by interaction between or merging of the two
jets \citep{Murphy2005,Murphy2008}. A single outflow or jet from a
binary system could also be produced by a single circumbinary disk
\citep{Machida2009,Mundt2010}. \citet{Reipurth2000} has also
postulated that the dynamical decay of multiple systems induces
outflow activity based on the fact that the binary frequency
(79\%-86\%) in 14 giant Herbig-Haro objects is twice higher than the
higher multiple frequency. 

At near-infrared (NIR) wavelengths, \FeII\ and H$_2$ emission lines are
excellent tracers of outflows and jets
\citep[e.g.][]{Nisini2002,Davis2001,Davis2003,Takami2006,Garcia2008,Garcia2010,Hayashi2009}.
Observations obtained with high-angular and high-velocity resolution,
particularly those that utilize large ground-base telescopes combined
with adaptive optics systems, allow us to study the detailed spatial
and kinematical structure close to the launching region
\citep{Pyo2003,Pyo2006,Takami2007,Beck2008,Davis2011}.  The \HeI\ $\lambda
$~1.083 $\mu$m line is a good tracer of the inner hot wind and funnel
flow.  These are evident as blueshifted absorption combined with
emission, and redshifted absorption, respectively
\citep{Edwards2003,Dupree2005,Edwards2006,Kwan2007,Fischer2008,Kwan2011}.
\citet{Takami2002} found that the \HeI\ emission associated with DG~Tau was
spatially extended toward the high-velocity blushifted jet direction.
\citet{Pyo2014} will present the detail spatial structure of this
\HeI\ extension.  Finally, the NIR {\ion{H}{1}} emission lines traces
accretion but also outflow activity.  For example, spectro-astrometry indicates that
the high velocity blueshifted gas observed in
Pa$\beta$ emission in DG~Tau is offset along the jet
direction \citep{Whelan2004}, while
\citet{Beck2010} have detected Br$\gamma$ emission
extended by more than 0$\farcs$1 from the star along the jet axes in
four CTTSs.

In this paper we present NIR 1 $\mu$m spectroscopy over a wavelength range
that covers \HeI\ $\lambda$~1.083 $\mu$m, \PaG\ $\lambda$~1.094
$\mu$m, and \FeII\ $\lambda~$1.257 $\mu$m emission lines obtained with the
integral field spectrograph, NIFS, at the Gemini North Observatory.
We focus on high-resolution \FeII\ emssion maps which show the complicated 
structure associated with the binary system UY Aur.

\section{OBSERVATIONS AND DATA REDUCTION}\label{sec:obs}

The data presented in this study were acquired on 2007 February 13
(UT) at the ``Fredrick C. Gillett'' Gemini-North 8m telescope located
on the summit of Mauna Kea, Hawaii, through the Subaru-Gemini time
exchange program under the Gemini program ID GN-2007A-Q-3.  We used
the AO-fed NIR integral field spectrograph NIFS \citep{McGregor2003}.
A modified J-band spectral setting, with central wavelength of 1.17
$\mu$m, was used to acquire spectra in the 1.06\,--1.28 $\mu$m range.
We chose this setting in order to simultaneously sample the emission
features of \HeI\ $\lambda$~1.0833 $\mu$m, \PaG\ $\lambda$~1.094
$\mu$m, and \FeII\ $\lambda$~1.257 $\mu$m.  These lines are known to
arise from accretion shocks, outflows, and jets, respectively, around
T Tauri stars.  The integral field unit (IFU) provided a spectral
resolution of R~$\sim$~5000 covering a 3$''\times$3$''$ field of view
with a pixel scale of 0$\farcs$1$\times$0$\farcs$04.  The NIFS data
were taken during photometric weather with an average natural seeing
of 0$\farcs$6--0$\farcs$85.  Altair, the Gemini-North facility AO
system, operates at 200~Hz at 700~nm with an optical stellar image as
the wavefront reference star.  We used UY Aur A as the reference star
because it was sufficiently bright in the R-band (R~$\sim$~11.4).  The
total on-source integration time was 1680 s ($=$ 120 s $\times$ 14
frames).  The position angle was set to 40$\degree$, which is the
orientation of the blueshifted micro-jet associated with this system
\citep{Hirth1997}.

The individual NIFS IFU data exposures were sky subtracted, flat
fielded, cleaned for bad pixels and rectified onto regularly sampled
spatial and spectral pixel grids using the processing tasks in the
Gemini IRAF package.  The standard NIFS data reduction process is
described in detail in \citet{Beck2008}.  Removal of telluric
absorption features was done using calibration observations of the A0
star HIP 23226 (J~$=$~8.029, 2MASS).  Removal of the Pa$\gamma ~ \lambda$~
1.094 $\mu$m absorption from the A0 standard star was done by fitting
Voigt profile shapes to the absorption, and interpolating and removing
any residual flux errors.  In the {\bf nifcube} data reduction task
used to generate 3-D datacubes from the NIFS data, we resampled the
data into a square grid of 0$\farcs$04 $\times$ 0$\farcs$04 pixels in
the spatial dimension.  The individual IFU datacubes were then shifted
so that the PSFs lined up at a central spatial location in the field,
and the multiple datacubes were combined using a median filter.  The
few frames that had measured seeing of more than 0$\farcs$85 were
excluded in the combining process, to improve the spatial sensitivity
of the final data.  The final combined cubes have a FWHM on the
stellar PSF of 0$\farcs$14 at the central wavelength of 1.17 $\mu$m.
IDL was used for all post-processing of the reduced data cubes: 
extraction of 1-D spectra, 2-D channel maps, position-velocity maps,
etc.

\section{RESULTS}\label{sec:results}

\subsection{\HeI , \FeII , and Pa$\gamma$ Line Profiles}\label{sec:spectra}
\begin{figure}
\epsscale{1.0}
\plotone{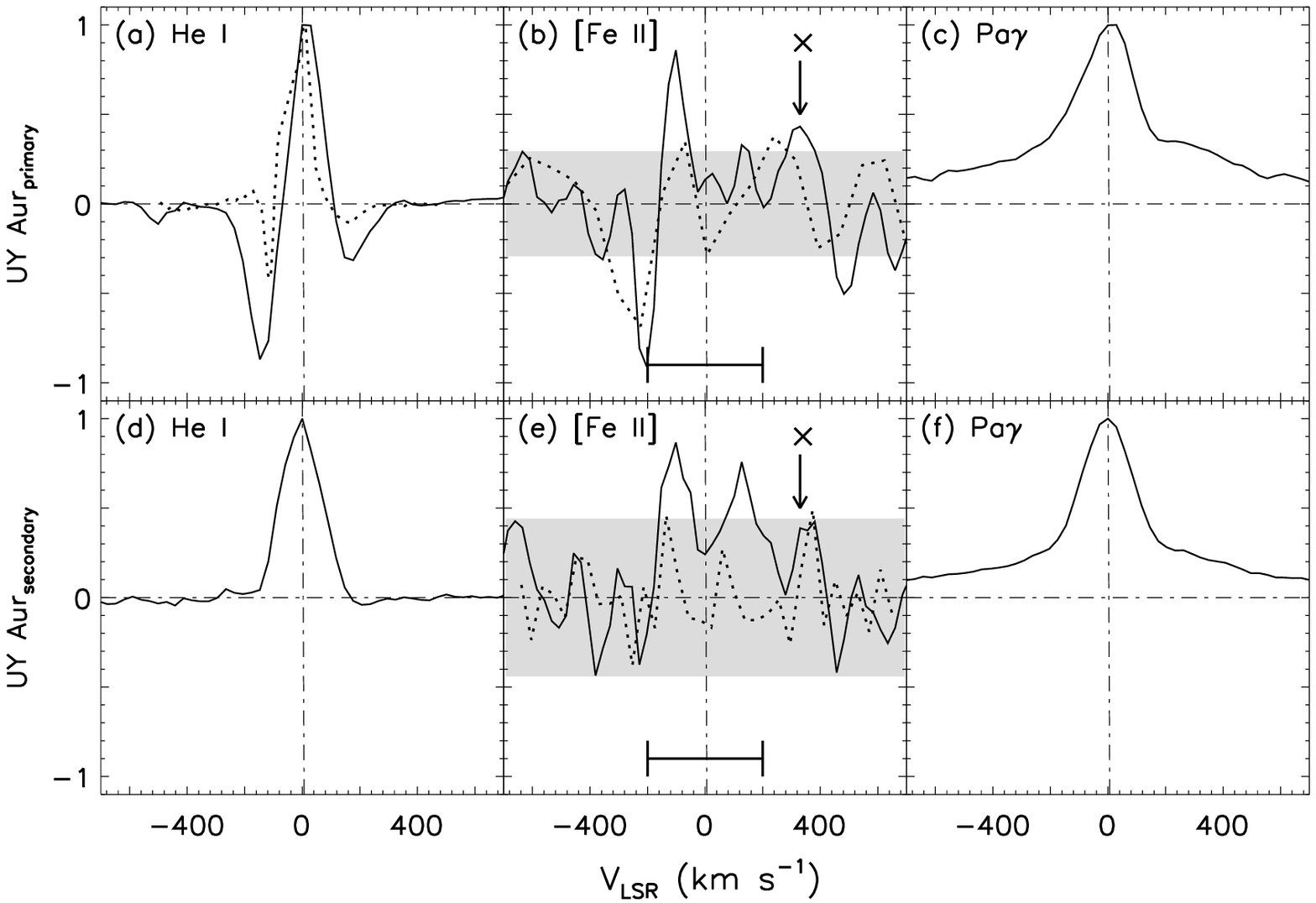}
\caption{Normalized line profiles toward UY Aur A (\textit{upper panels}) 
 and UY Aur B (\textit{lower panels}) created by integration of the flux within 0$\farcs$4 diameter aperture:
 (\textit{left}) \HeI\ $\lambda$ 1.0833 $\mu$m. The dotted line in panel $\textit{a}$
is \HeI\ line profile observed by \citet{Edwards2006} in November 2002. 
(\textit{middle}) \FeII\ $\lambda$ 1.2570 $\mu$m. The dotted lines are normalized photospheric
spectrum of M0V type \citep[HD 19305,~R~$=$~2000: ][]{Rayner2009} in $\textit{b}$ and M2V type 
\citep[GL 411,~R~$=$~3000: ][]{Wallace2000} in $\textit{e}$, respectively. They are normalized with the the peak flux at the point marked with $\times$. 
  (\textit{right}) \PaG\ $\lambda$ 1.0941 $\mu$m.
The dot-dashed vertical lines at V$_{\rm LSR}=+$6.2 \kms\ indicate the
  systemic velocity of UY~Aur.  The horizontal bars located at the
  bottom of panels $\textit{b}$ and $\textit{e}$ mark the defined
  \FeII\ emission line ranges; the shaded areas in these two panels show
  the range where the photospheric lines are dominant; the emission
  features marked with an $\times$ in these panels are not \FeII\ emission.
\label{fig_profile}}
\end{figure}

Figure~\ref{fig_profile} shows the normalized line profiles of
\HeI\ $\lambda$~1.0833 $\mu$m, \FeII\ $\lambda$~1.257 $\mu$m, and
Pa$\gamma$ $\lambda$~1.094 $\mu$m emission from UY Aur A (primary) and
B (secondary).  To extract these spectra we integrated the flux within
a 0$\farcs$4 diameter ($\sim$~3\,-- 4 times the FWHM of the achieved
spatial resolution) around each star.  The systemic velocity of UY~Aur
is V$_{LSR}~\sim$~6.2 \kms\ \citep{Duvert1998}. The system has an
orbital period of $\sim$~2074 yr and the orbital maximum velocity is
3.13~$\pm$~0.37 \kms\ \citep{Close1998}. We use the LSR velocity system 
for all velocities quoted in this paper.

The \HeI\ $\lambda$~1.0833 $\mu$m line profiles observed towards UY Aur
A and B are remarkably different.  The \HeI\ line towards UY Aur A
possesses deep blue-shifted and redshifted absorption features (85.5
\% and 33.5 \% of the emission, respectively).  The total amount of
absorption flux is 19 \% larger than that of emission flux.  Indeed, a
pure \HeI\ line image made by integrating the flux density from $-$600
\kms\ to $+$600 \kms\ has a negative net flux.  The deepest trough in
the blueshifted absorption feature is located at $-$150 \kms ; the
wings of this absorption feature reach a maximum velocity of $-$400
\kms .  The deepest point in the redshifted absorption feature is at
$+$170 \kms ; the wings of this absorption feature reach $+$300 \kms .
Notably, the \HeI\ profile shows more developed absorption dips than
were observed by \citet{Edwards2006} in November 2002.  This is
perhaps not surprising since, as noted by Edwards et al., 
\HeI\ profiles can vary, even over a two night interval.  In stark
contrast to UY Aur A, the \HeI\ profile observed towards UY Aur B has
a very simple, symmetric emission line profile that is confined to
within $\pm$200 \kms. No blueshifted nor redshifted absorption is
seen.  The peak flux of \HeI\ emission from UY Aur B is $~$25 times
smaller than that of UY Aur A.

The \FeII\ $\lambda$~1.257 $\mu$m line profiles from UY Aur A and B
are shown in Figures~\ref{fig_profile}$\textit{b}$ and
\ref{fig_profile}$\textit{e}$, where the velocity ranges of the
\FeII\ emission are shown by the horizontal bars.  These \FeII\ lines
were detected at greater than 3$\sigma$ of the background noise from
both stars, but are marginally seen in these spectra where the
continuum emission is strong and many photospheric features are seen.
The primary and secondary both have emission peaks at velocities at
$-$100 \kms\ and $+$130 \kms\ .  The absorption and emission features
at $-$200 \kms\ and $+$300 \kms, respectively, are photospheric
features from the star.
The dotted lines represent the spectra of M0V and M2V type stars, 
objects that have spectral types similar to UY Aur A and B, respectively. 
The spectral resolution of these standard star observations is lower than 
our data. Even so, they demonstrate how the \FeII\ emission within $\pm$200~\kms 
~is affected by photospheric features. The blueshifted \FeII\ emission peak 
of the primary is two times higher than the photospheric feature, but the 
\FeII\ emission of the secondary is dominated by both blue-shifted and 
red-shifted photospheric features. 

\PaG\ emission is a surrogate to trace magnetospheric accretion onto
the star. The profiles observed towards UY Aur A and B in
Figure~\ref{fig_profile}$\textit{c}$ and \ref{fig_profile}$\textit{f}$
possess well-developed wings.  Broad {\ion{H}{1}} emission with no
absorption is typical for a magnetospheric origin.

\subsection{\FeII\ and Continuum Images}\label{sec:images}

\begin{figure}
\epsscale{0.8}
\plotone{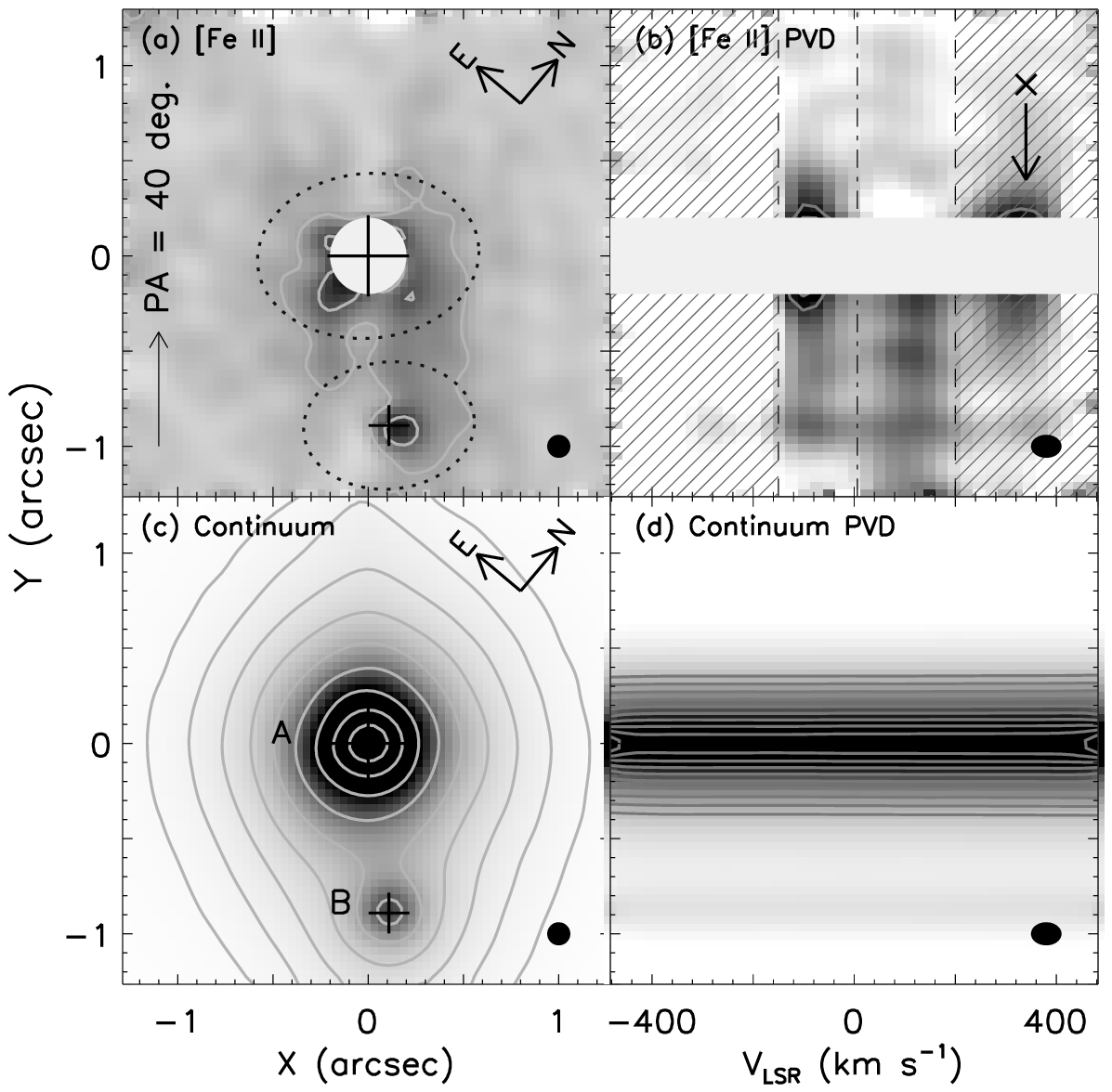}
\caption{(\textit{a}) Monochromatic \FeII\ $\lambda$~1.257 $\mu$m
  image integrated across the residual flux from $-$150 \kms\ to
  $+$200 \kms\ .  This image includes the weak absorption feature from
  $-$60 to $+$100 \kms\ within Y~$=~\pm$0$\farcs$2.  The dotted ellipses
  surrounding the two stars shows the projected effective Roche lobe
  \textit{r$_L$} \citep{Eggleton1983}.  (\textit{b})
  \FeII\ Position-Velocity diagram with the flux integrated along the
  X-axis direction in our data cube.  
  We have masked the area  within 0$\farcs$2\ in radius of the primary star with white patches. Here residuals associated with the continuum subtraction obscure our view of this region.  
  The feature labelled with an $\times$ is a photospheric line.  
The dot-dashed vertical
  line indicates the systemic velocity of UY~Aur  (V$_{\rm
    LSR}=+$6.2 \kms\ ).  (\textit{c}) Continuum image and (\textit{d})
  PV diagram showing the continuum range.  The large plus mark in
  (\textit{c}) indicates the peak position of UY Aur A; the small plus
  is centered on UY Aur B.  The Y axis in each panel is aligned with
  an on-sky position angle of 40$\degree$.  The resolution of each
  panel is displayed with an oval bottom-right. In these data the
  spatial resolution is 0$\farcs$14 and the velocity resolution is 60
  \kms .
\label{fig_PVD}}
\end{figure}

Monochromatic images produced by integrating across the \HeI\ and
\PaG\ emission lines do not show any spatial extension beyond a simple
point source distribution. Thus we do not discuss these data further
in this section and instead concentrate on our \FeII\ data.  A
continuum subtracted \FeII\ line image is presented in
Figure~\ref{fig_PVD}$\textit{a}$; a continuum image is shown in
Figure~\ref{fig_PVD}$\textit{c}$.  To investigate the structure of the
gas emission in the context of the gravitational influence of the two
stars, we have calculated the individual Roche lobe radii for UY Aur A
and B.  We use this Roche lobe analysis as an approximation for the
sphere of influence of each star, to better understand the structure
of the gas emission.

The \FeII\ line image in Figure~\ref{fig_PVD}$\textit{a}$ was
extracted from the reduced IFU data cube by integrating over the
velocity range $-$150 to $+$200 \kms . Note that by choosing this
range we exclude the deep stellar absorption feature adjacent to the
blueshifted \FeII\ emission peak (see
Figure~\ref{fig_profile}$\textit{a}$).
Figure~\ref{fig_PVD}$\textit{b}$ shows a position velocity diagram for
\FeII\ plotted along the Y-axis in Figure~\ref{fig_PVD}$\textit{a}$,
with the emission integrated across the entire X-axis direction.  In
Figure~\ref{fig_PVD}$\textit{a}$ the effective Roche
lobes\footnote{The effective Roche lobe radius is defined as:
  $\textit{r} _{L_1} = \frac{0.49q^{\frac{2}{3}} A}{0.6 q ^\frac{2}{3}
    + ln( 1 + q ^\frac{1}{3} )}$, where $A=$~length of semi-major
  axis, $q~= M_1/M_2 $ \citep{Eggleton1983}.} associated with UY Aur A
and B are indicated by ellipses drawn with a dashed line. In defining
these areas we have assumed that $q=$1.765 \citep[$= M_1/M_2 =
  0.60/0.34$;][]{HK2003} and $A=$ 190 AU \citep{Close1998}.  The
\FeII\ emission in Figure~\ref{fig_PVD}$\textit{a}$ appears to fill
the Roche lobe around the primary, UY Aur A, although the emission is
clumpy. The emission around the primary also extends toward the SW
where it connects to a faint peak associated with the secondary
source, UY Aur B. There is no emission to the south of the secondary.

The continuum image was made by integrating the spectra over 
$\pm$22.458 \AA\ around 1.2250 $\mu$m and is shown in Figure~\ref{fig_PVD}$\textit{d}$.  The
large and small '$+$' symbols in \ref{fig_PVD}$\textit{c}$ indicate
the positions of the primary and secondary sources, respectively.  The
continuum image (\ref{fig_PVD}$\textit{c}$) clearly resolves the
0$\farcs$897 separation between UY Aur A and B.  
The position-angle of the axis that links UY Aur A and B is 226$\dotdeg$97; the position angle of the semi-major axis of the circumbinary disk that surrounds UY Aur A and B is 135$\degree$.
The two stars appear to lie within the plane of the circumbinary disk \citep{Close1998}.
The inclination angle of the disk with respect to the line of sight is 42$\degree~\pm$~3$\degree$ \citep{Hioki2007,Close1998}.  After correcting for this inclination angle, the separation between A and B is 1$\farcs$21$^{+0.00} _{-0.05}$ and the position angle of the axis between the two sources is 226$\dotdeg$46~$\pm$~0$\dotdeg$07.
The separation is 0$\farcs$02 larger than was measured by
\cite{Hioki2007} and \citet{Close1998}.  The position angle is 2$-$3
degrees smaller than values cited by  \citet{Hioki2007}.  The
continuum flux ratio of the primary (A) with respect to the secondary (B) is 15.4 $\pm$ 3.9  ($=$ 2.97$^m \pm$ 1.5$^m$ ); \cite{Close1998} measured in 1996 a J-band flux
ratio of 2.19 mag.

\subsection{\FeII\ Velocity Structure}\label{sec:v_structure}

Figure~\ref{fig_PVD}$\textit{b}$ shows a position-velocity diagram
(PVD) for \FeII\ over the range $\pm$500 \kms.  
\citet{Hirth1997} reported that they detected high-velocity blueshifted gas with 
a radial velocity of $-$200 to $-$100 \kms\ in \SII\ in 1992 December although they did not present the data. 
They could not spatially resolve the binary because the data was limited by the seeing of 1$\farcs$4 -- 2$\farcs$0.   
Our high spatial resolution data clearly show that the blueshifted \FeII\ emission
between $-$150 and $-$50 \kms\ is strongest around the primary star
and extends toward the secondary star. 
The redshifted \FeII\ emission between $+$50 and $+$200 \kms\ is evident at Y $<-$0$\farcs$1.

\begin{figure}
\epsscale{1.0}
\plotone{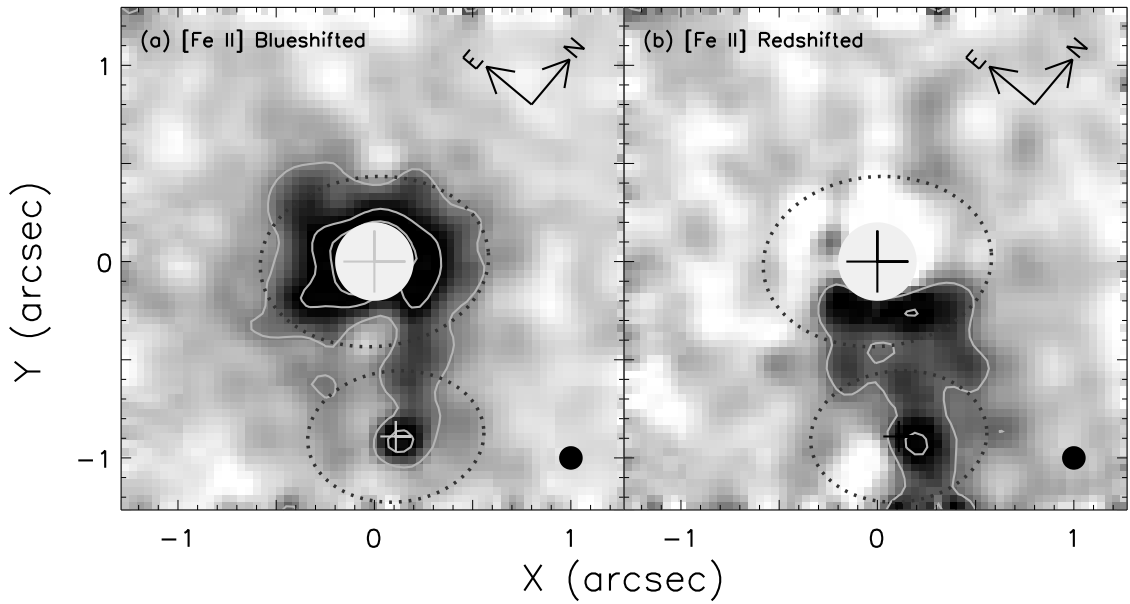}
\caption{(\textit{a}) Blueshifted \FeII\ emission integrated over the range
  $-150< \VLSR <-60$ \kms .  This range excludes the ambient
  emission between $-$60 and 0 \kms\ .  (\textit{b}) Redshifted
  \FeII\ emission integrated over the range
  $+60< \VLSR <+200$ \kms .  The dotted ellipses show the effective Roche lobe
  radii projected on the sky.  The oval in the lower-right corner of each panel shows the
  spatial resolution, as
  described in Figure~\ref{fig_PVD}.
\label{fig_RB}}
\end{figure}

In Figure~\ref{fig_RB}$\textit{a}$ and \ref{fig_RB}$\textit{b}$ we
present images of the blueshifted and redshifted \FeII\ emission,
respectively.  The blueshifted emission fills the inside of the Roche
lobe surrounding the primary, while the redshifted emission is widely
spread to the south-west of the primary with 0$\farcs$2 offset at Y $<-$0$\farcs$1
in Figure~\ref{fig_RB}$\textit{b}$.
A portion of the blue-shifted and red-shifted gas fills a 'bridge'
that links the western side of the primary with the secondary (along
X~$\sim$~0$\farcs$1 -- 0$\farcs$4).  Inside the Roche lobe of the
secondary, the blueshifted and redshifted emission occupies only a
narrow ridge that runs along the NW side of the source connecting up
with the gas SW of the primary.  The redshifted emission near the
secondary is more extended toward the SW direction (PA
~$=$~220$\degree$).  The SE side of the secondary does not exhibit any
emission at all.  The blueshifted emission peaks are
offset 0$\farcs$1 NE of the primary and 0$\farcs$04 W of the
secondary;  the redshifted emission peak near the secondary is
offset 0$\farcs$1 to the NW.  The blueshifted emission is a few times
brighter than the redshifted peak.
  
\begin{figure}
\epsscale{1.0}
\plotone{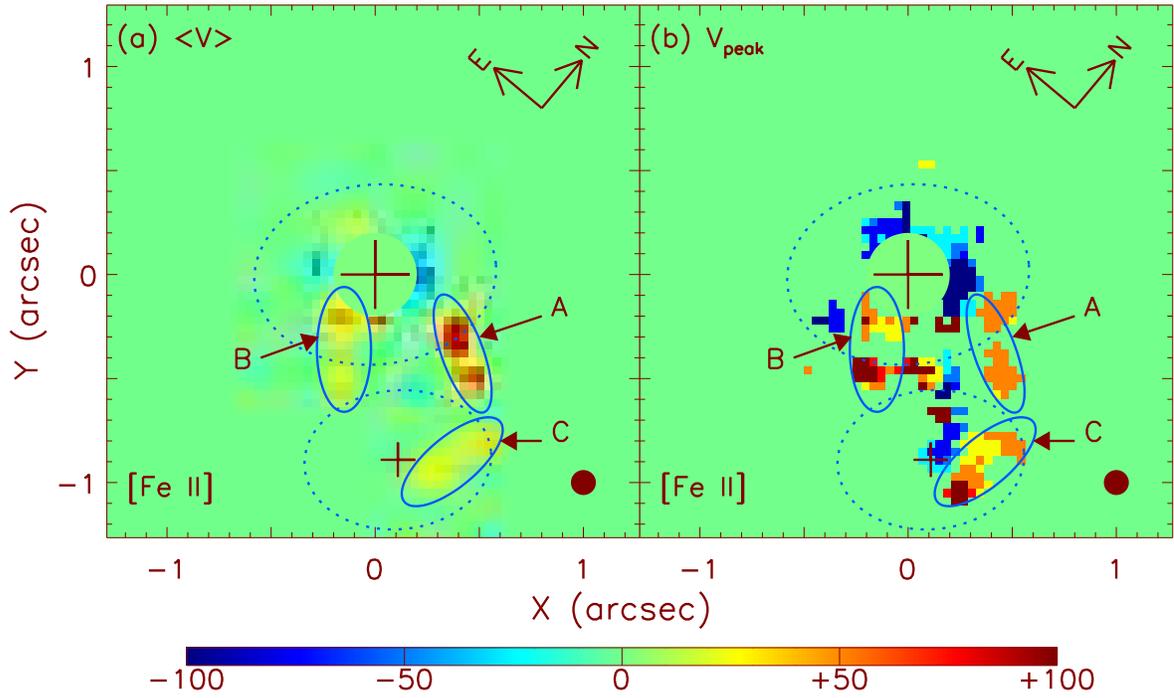}
\caption{(\textit{a}) Velocity barycenter (flux weighted mean
  velocity: $< V >$) of the \FeII\ emission; (\textit{b}) Velocity at peak
  flux (V$_{peak}$).  The blue dotted ellipses show the effective
  Roche lobe radii projected on the sky.  The blue solid ellipses mark
  the regions filled with redshifted gases.  The oval in the lower-right
  corner of each panel shows the
  spatial resolution, as described in Figure~\ref{fig_PVD}.
\label{fig_Vpeak}}
\end{figure}

In Figure~\ref{fig_Vpeak}$\textit{a}$ and \ref{fig_Vpeak}$\textit{b}$
we plot the intensity weighted mean velocity or velocity
barycenter\footnote{$< V >~ = ~ \frac{\int v I_v \mathrm{d}v}{\int I_v
    \mathrm{d}v}$ is the velocity centroid, where $I_v$ is the intensity of emission at
  velocity $v$) \citep{Beck2008,Riera2003}.}, $<V>$, as well as the
velocity of the emission peak\footnote{$V_{\rm peak}$ is defined as the
  velocity with peak intensity MAX$(I_v)$.}, $V_{\rm peak}$.  The velocity
range for the calculation in each case is from $-$150 \kms\ to $+$200
\kms.  $<V>$ represents the effective velocity moment of the
emission line profile at each point, while $V_{\rm peak}$ is useful for
distinguishing the flow tendency.  In
Figure~\ref{fig_Vpeak}$\textit{a}$, the effective blueshifted gas
shows a `V' shape morphology with a 90$\degree$ opening angle
pointing upward (PA~$\sim$~40$\degree$) from a point Y~$=-$0$\farcs$1
below the primary.  The effective redshifted gas is pronounced to the
west (region `A') and SW (region `B') of the primary, and to the NW
(region `C') of the secondary. 
The `bridge' structure shown in Figure~\ref{fig_RB}$\textit{a}$
 along the Y direction at X $\sim$~0$\farcs$1 -- 0$\farcs$4 disappears in the
$<V>$ map. This indicates that blueshifted and redshifted
gases co-exist in this area, together making $<V>~\sim$~0 \kms.
Overall, the effectivelly redshifted gas
shows a clumpy structure.  The most redshifted clump is in region `A'
at (X, Y)~$=$~(0$\farcs$38, $-$0$\farcs$3).  This feature is elongated
along a PA~$=$~240$\degree$ with respect to the primary and represents
fast, redshifted gas with $<V>~\sim+$100 \kms .  The region `B' ($<V>~\sim+$25 \kms ) is located at
(X, Y)~$=$~($-$0$\farcs$15, $-$0$\farcs$36) and is elongated along a 
PA~$=$~220$\degree$ with respect to the primary. The regions `A' and `B'
correspond to the boundary seen in Figure~\ref{fig_RB}$\textit{b}$.
Region `C' ($<V>~\sim~+$25 \kms ) is located to the west of the
secondary at (X, Y)~$=$~(0$\farcs$28, $-$0$\farcs$9) and is elongated
along a PA~$=~-$10$\degree$.

The distribution of velocities at the emission line peak, $V_{\rm peak}$
(Figure~\ref{fig_Vpeak}$\textit{b}$), is somewhat different from the
$<V>$ map.  
The gas with blueshifted $V_{\rm peak}$ is spread around the primary at Y $\ga$~0$\farcs$2. 
The `bridge' of blueshifted gas reappeared here along the Y direction elongated from the vicinity of the secondary to the NW of the primary as was seen in Figure~\ref{fig_RB}.  
The fastest blueshifted emission peak, $V_{\rm peak}$, is located at (X, Y)~$=$~(0$\farcs$25, $-$0$\farcs$05), at the end of the extension.
Most of the gas clumps with redshifted $V_{\rm peak}$ are located within the regions identified as `A', `B', and `C'.  
The fast ($\sim~+$100 \kms) redshifted clumps are found at (X, Y)~$=$~($-$0$\farcs$2, $-$0.45) in `B' region, (0$\farcs$25, $-$1$\farcs$05) in `C' region. 

\section{DISCUSSION}\label{sec:discussion}

The primary shows deep blueshifted absorption in the
\HeI\ $\lambda$~10830 line profile (Figure~\ref{fig_profile}).
\citet{Edwards2003,Edwards2006} pointed out that the diversity in width
and depth of the blueshifted absorption of \HeI\ indicates an inner
wind emanating from a star with a large solid angle.  \citet{Kwan2007}
classified the \HeI\ line profile they observed towards UY Aur as a
disk wind with the second source of emission because the profile
in \citet{Edwards2006} showed narrow blueshifted absorption and
emission extended to blueward. However, the profile which we observed
on 2007 (Figure~\ref{fig_profile}$\textit{a}$) showed more developed
blueshifted absorption and resembled the profile of CY~Tau in \citet{Kwan2007}, 
which were interpreted in terms of a stellar wind model.  

The \PaG\ $\lambda$~1.094 $\mu$m emission does not show any spatial extension.
This is consistent with the interpretation that the Br$\gamma$ and \PaG\ emissions arise from very compact magnetospheric accretion columns in the vicinity of the central stars. 
On the other hand, \citet{Beck2010} reported that 50\,\% of their targets (four out of eight) had spatially extended Br$\gamma$ emission, suggesting that the atomic Hydrogen emission can originate from extended jets or scattering by extended dust around the central stars.
They assumed that the other four objects also had spatially extended Br$\gamma$ emission that was obscured by the bright continuum emission close to the central stars. 
The extended emission could not have been detected if the physical conditions of the jet gas were not adequate: i.e., if the temperature were too low or the density were too high. 
In the case of spatially extended dust scattering, the spatial distribution of the line emission should follow that of the continuum emission.

The velocity structure between UY Aur A and B is complicated because of the overlap of blueshifted and redshifted emission features.
A narrow structure connecting the primary and secondary disks similar to the `bridge' observed in the present studies is predicted by accretion flow simulations of circumbinary disks \citep{Gunther2002,Hanawa2010, Fateeva2011}. 
The bridge might then be related to the boundary region between two colliding accretion flows. 
If the circumbinary disk planes are not perpendicular to the line of sight, both blueshifted and redshifted accretion flows would exist around the `bridge' structure.
However, it is difficult to produce the high velocity features seen in our data as a result of colliding flows between the two circumstellar disks because the accretion  gas flows should basically have Keplerian velocities, of the order of a few \kms\ at the edge of circumstellar disks. 
Thus, the high velocity emission features are probably associated with outflows launched from the vicinity of the central stars. 

The \FeII\ emission profile observed towards the UY Aur system
exhibits blueshifted emission at $-$154 -- $-$46 \kms\ and redshifted
emission at $+$67 -- $+$201 \kms\  (after correcting for the 42$\degree$
inclination angle of the circumbinary disk). 
Because the terminal velocity of an outflow is comparable to the 
Keplerian rotation velocity of a disk at the launching region
\citep{Kudoh1998}, the fast flow velocities indicate that the launching 
region is located in a deep potential well, close to the star in the accreting star-disk system.
If we use the stellar masses of the primary and secondary as 0.6\,$M_\sun$ and 0.34\,$M_\sun$, respectively, estimated by \citet{HK2003} using the pre-main-sequence tracks of \citet{Siess2000}, and assume three-times larger lever arm radii than the foot point radii for magnetocentrifugal acceleration  \citep{Konigl2000},
then the foot points for outflow acceleration correspond to $\sim$~0.22\,--\, 0.6\,AU and $\sim$~0.12\,--\,0.34\,AU for UY Aur A and B, respectively.
The other estimates of stellar masses are more or less consistent with that of \citet{HK2003}.
Based on the Keplerian rotation measured by the $^{13}$CO gas in the circumbinary
disk, \citet{Duvert1998} estimated the total mass of the UY Aur binary
as $\sim$~1.2 $M_\sun$.
\citet{Close1998} estimated the total mass of
the binary as 1.61$^{+0.47}_{-0.67}$ $M_\sun$ using orbital parameters
derived from observations obtained since 1944.
Together these
results show the uncertainty of the total mass estimate of the binary system, not affecting the above estimates of launching radii.

The geometrical distribution of \FeII\ emission from the primary
suggests a wind with a wide opening angle: the blueshifted and
redshifted emission in Figure~\ref{fig_RB} is widely distributed, and
in the diagram of $<V>$ (Figure~\ref{fig_Vpeak}$\textit{a}$) the blueshifted emission
opens out in a `V' shape around the primary. 
The gap of 0$\farcs$2 between the primary and the redshifted emission 
can be explained by the optically thick circumstellar disk of UY Aur A.
The FWHM diameter of the
star in our images was measured to be $\sim$~0$\farcs$14. We have to
be careful when interpreting structures within $\sim$~0$\farcs$1 of the
star because the subtraction residuals associated with Poission noise
in the central region can hugely affect the structure. However, we can
say that the blueshifted gas is spread over more than $\sim$~0$\farcs$2
radius.  We note that a similar fast wind with a wide opening angle
has been reported from at least one other young source, L1551 IRS 5
\citep{Pyo2005,Pyo2009}.

The complicated velocity structure in UY Aur prevents us from easily identifying the origin of the various emission features. The high-velocity redshifted features labelled A and B in Figure~\ref{fig_Vpeak} could be assocaited with the edges of a fast wind with a wide opening angle from UY Aur A. 
 Feature `C' could then be an extension of feature `A'.  
With this model, however, it is diffcult to explain the blueshifted emission seen between features A, B and C. This emission may be associated with a micro jet from the secondary, UY Aur B. Clearly, deeper observations with spatial and spectral resolutions comparable to those presented here are required in \FeII\ to further probe this complex region. 
The elongated knot of redshifted emission labelled region `C'
in Figure~\ref{fig_Vpeak} may be associated with the primary rather
than the secondary, i.e. it may be an extension of region `A'.  The
secondary may drive only a narrow blueshifted jet towards the North.
Deeper observations with spatial and spectral resolutions
comparable to the presented studies are required in the \FeII\ line to further
probe this complex region.

The absence of \FeII\ emission at the south of the secondary (within its Roche lobe) in Figure~\ref{fig_RB}{\em b} is curious. 
Intrinsically existing emission could have been obscured by a gas-dust cloud ejected from the binary or the secondary.  
A notably large flux variation of the secondary may be caused buy such a cloud.
For example, in 1992 it became 5 mag fainter than the primary in the R-band \citep{Herbst1995}. 
The R-band magnitude difference ($\Delta$m) between the primary and secondary has changed by 6.6$^m$ in 1996 \citep{Close1998} and 2.0$^m$ in 2000 \citep{Brandeker2003}. 
In the H-band, the $\Delta$m changed 1.7$^m$ in 1996, 0.43$^m$ in 2002, and 1.7$^m$ in 2005. 
Overall, the secondary has exhibited larger magnitude variations than the primary: $\pm$1.3$^m$ compared to $\pm$0.3$^m$ between 1996 - 2005 \citep{Hioki2007, Close1998}.
These magnitude variations should be related to the changes in mass loss (outflow) and mass accretion rates. 
\citet{Berdnikov2010} pointed out that a gas-dust cloud had obscured UY Aur during 1945 - 1974. 
This cloud may still be responsible to obscure UY Aur B.

\citet{Skemer2010} reported that the silicate spectrum of UY Aur B is much flatter than expected. 
This may be the result of foreground extinction for an edge-on disk: the UY Aur disk may be viewed edge-on, although such a disk has not been detected in direct imaging observations of UY Aur B.
Thus the disk may not be precisely `edge-on', i.e., the inclination may be larger than $\sim$ 15 degrees with respect to the line of sight.
\citet{Monin2007} postulated that the mis-alignment of circumstellar disks in binary systems may be quite common in wide binaries (with a $>$ 100 AU) for Class 0, I , and II phases. 
UY Aur system may be an example of such a case.
    
\begin{figure}
\epsscale{1.0}
\plotone{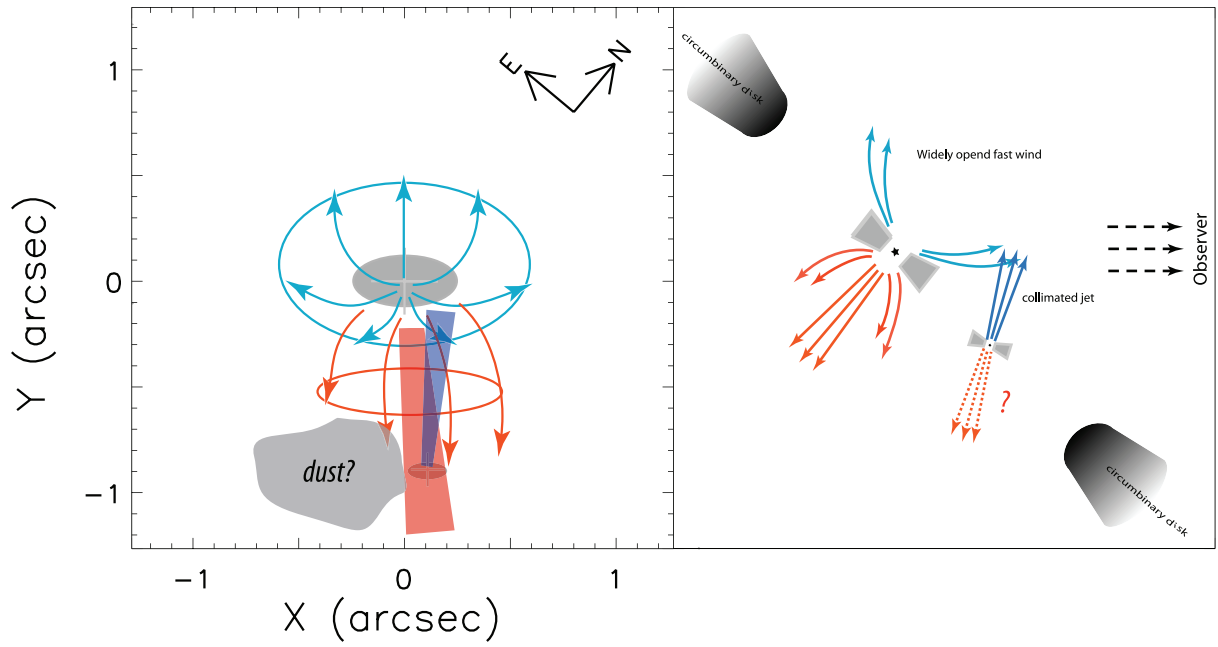}
\caption{(\textit{left}) Schematic drawing of the redshifted (\textit{red}) and and blueshifted (\textit{blue}) outflows emanating from UY Aur A and B on observed image plane. 
(\textit{right}) Schematic view of the model of the UY Aur binary and their outflow and jet.
\label{fig_draw}}
\end{figure}

Figure~\ref{fig_draw} shows a schematic drawing of the UY Aur system, illustrating our overall understanding of this remarkable object.

\section{SUMMARY}

Our high spatial and high velocity resolutions NIR integral field
spectroscopy of the UY Aur young interacting binary system show the
following results:

\begin {enumerate}

\item{We detected \HeI\ $\lambda$~1.0830 $\mu$m emission from both the
  primary and the secondary components of the binary, UY Aur A and B. UY
  Aur A shows deep blueshifted and redshifted absorptions with a
  central emission line which supports the stellar wind model
  interpretation of \citet{Kwan2007}.  UY Aur B shows only emission,
  25 times fainter than that of the primary.}

\item{Our observations suggest that UY Aur A drives fast, blueshifted and redshifted \FeII\ outflows with wide opening angles.  Between binary components A and B we observe complex velocity structure with
overlapping blueshifted and redshifted emission features.  
We identified a gas `bridge' between the primary and secondary stars, although the high velocities observed along the edges of this feature support an outflow interpretation.  UY Aur B may then be associated with a blushifted micro-jet. 
}

\end {enumerate}

\acknowledgments We appreicate for the valuable of comments from the anonymous referee. 
This research has been made using NASA's Astrophysics
Data System and the SIMBAD database operated at CDS, Strasbourg,
France.


\clearpage

\clearpage

\clearpage

\clearpage

\clearpage

\end{document}